\documentclass[aps,prl,preprint,groupedaddress,showpacs]{revtex4-1}
\usepackage{amsmath,amsthm,amssymb,latexsym}
\usepackage{graphicx}
\usepackage{subfigure}  

\allowdisplaybreaks
\allowdisplaybreaks
\begin{document}

\title{Atoms ionized by assisted attopulses behaving as diatomic molecules}

\author{D I R Boll}
%\email[]{boll@ifir-conicet.gov.ar}
\affiliation{Instituto de F\'{\i}sica Rosario and Laboratorio de Colisiones At\'omicas, CONICET-UNR, Bvrd. 27 de Febrero 210 bis, 2000 Rosario, Argentina}

\author{O A Foj\'on}
%\email[]{fojon@ifir-conicet.gov.ar}
\affiliation{Instituto de F\'{\i}sica Rosario and Laboratorio de Colisiones At\'omicas, CONICET-UNR, Bvrd. 27 de Febrero 210 bis, 2000 Rosario, Argentina}

\altaffiliation{Escuela de Ciencias Exactas y Naturales, Universidad Nacional de Rosario, Argentina.}

\date{\today}

\begin{abstract}
The single ionization of noble gas atoms by the combined action of XUV attopulses and an infrared laser field is theoretically investigated by means of a non-perturbative model that under certain approximations gives closed-form expressions for the angular distributions of photoelectrons. Interestingly, our model allow us to interpret the angular distributions as two-center interferences where the separation between the centers is governed by the infrared laser field. Angular distributions are compared to the available experimental data showing a good agreement. Finally, we deduce the conditions to obtain zeros in the angular distributions coming from destructive two-center interferences.       
 
\end{abstract}


\maketitle
The development of tools able to scrutinize the electron dynamics in its own time scale attracted an increasing attention since the first realization of attosecond pulse trains \cite{Paul2001,Hentschel2001}. The subsequent diversification of techniques have built the realm of attophysics in which the coherent control of electron dynamics, in atoms or molecules, emerged as one of the most fascinating perspectives. Moreover, the control of electron localization in dissociating molecular states \cite{Kelkensberg2011} and the control of orbital parity mix \cite{Laurent2012} have been proved to be valid tools to steer dynamical properties in reactions.

When atomic or molecular targets are exposed to the  simultaneous action of an attosecond pulse train (ATPT) of odd harmonics and a low intensity near infrared laser (NIR) field, the reconstruction of attosecond beating by interference of two-photon transition scheme is obtained. The spectrum contains dressed harmonic (DH) lines mainly populated by the absorption of a given harmonic in the ATPT, and sideband (SB) lines associated to the further exchange of NIR photons. The magnitude of these SBs oscillates at twice the NIR frequency when the  ATPT-NIR delay is modified. This measurement scheme lies at the heart of the attosecond physics allowing the reconstruction of the ATPT time structure \cite{Paul2001} and the time delay determination in photoionization \cite{Dahlström2012}. The intermediate NIR intensity regime, where the exchange of more-than-one NIR photon is expected, has received much less attention.  The presence of many interfering quantum channels requires a treatment beyond the second-order perturbation theory \cite{Paul2001,Laurent2012}. 

The theoretical approach to these problems is by no means simple. Solving the Time Dependent Schr\"odinger Equation (TDSE) for reactions such as the photoionization of multi-electron atomic targets assisted by a NIR represents a computational challenge \cite{Galan2013} for the current computational resources. The use of simplified models leading to predictions in reasonable agreement with \emph{ab-initio} calculations and/or experimental results reveals as a valuable tool to understand the physical processes involved, as the numerical results do not often have a straightforward interpretation. Nowadays, models able to describe reactions assisted by a stronger NIR are available. Among them, the soft-photon approximation \cite{Maquet2007} was successfully applied to study angular distributions (ADs) in laser-assisted atomic photoionization by photons from free electron laser \cite{Meyer2008} or high harmonic generation (HHG) \cite{Galan2013,Picard2014} sources. Also, it was established in previous studies for atomic/molecular targets \cite{YPCB,YCB,Boll2014} that the Separable Coulomb-Volkov model (SCV) may provide results in qualitative agreement with \emph{ab-initio} calculations. 

The sophisticated techniques involved in the measurement of energy- and angle-resolved spectra have evolved in an outstanding way. The long-term stability of the ATPT-NIR synchronization, necessary for this kind of experiment in cold target recoil ion momentum spectroscopy devices, was reduced to about 60 attoseconds \cite{Picard2014,Weber2015}, enabling thus the determination of the AD of photoelectrons with a given energy and an almost fixed ATPT-NIR delay. Experimental studies concerned with the photoelectron ADs, although scarce, showed a critical dependence with the ATPT-NIR delay \cite{Guyetand2005,Remetter2006,Varju2006,Guyetand2008,Picard2014,Weber2015}. The global shape of the ADs in DH lines changes significantly for different delays, as opposed to the SB lines that, after normalization, are almost insensitive to the delay change. \cite{Picard2014}. In this context, the ADs pose a stringent test to the theoretical treatment.

\begin{figure}[t!]
\includegraphics[width=\columnwidth]{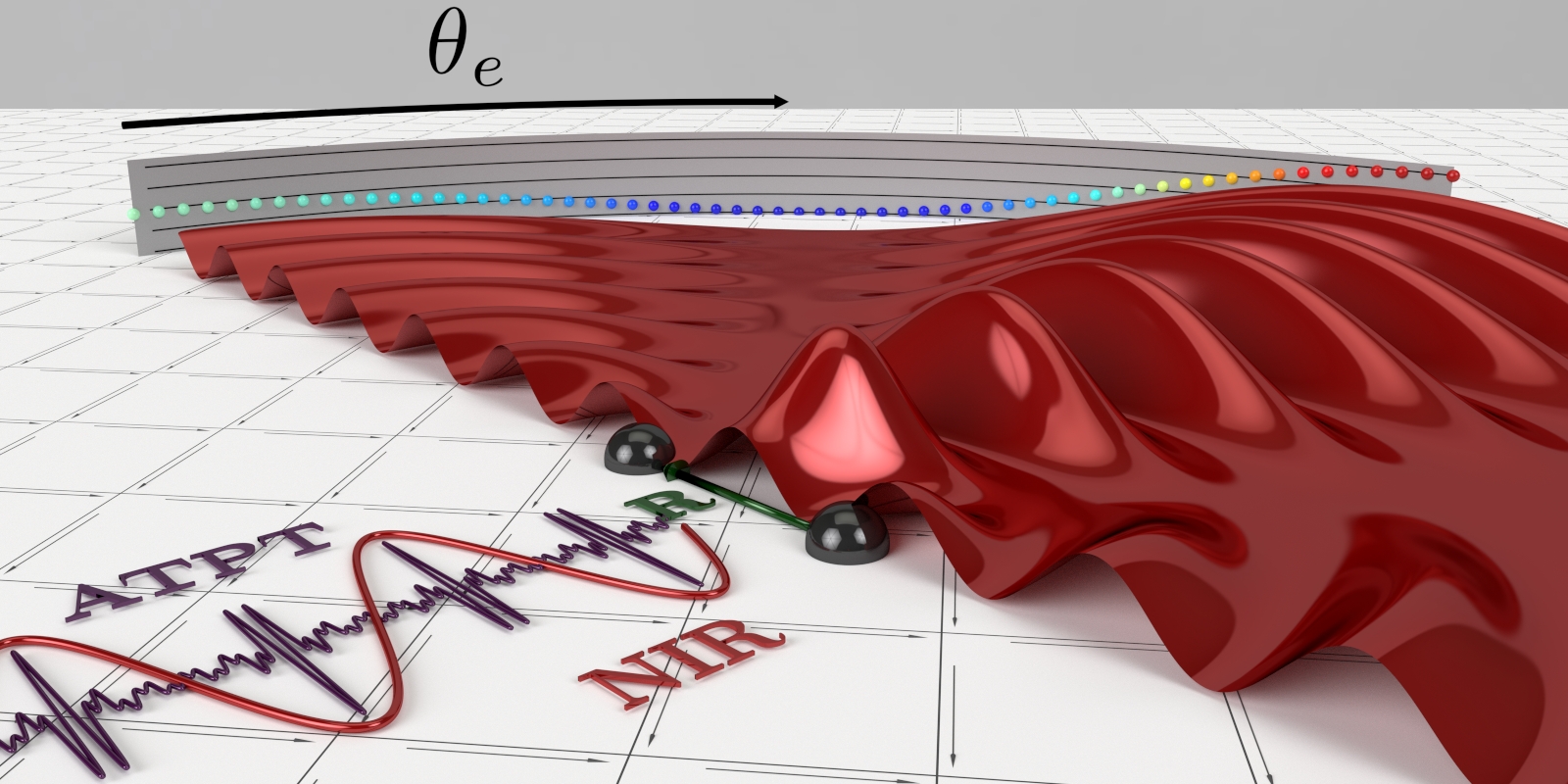}
\caption{\label{fig:illustration}(Color online)  Pictorial representation of the reaction of interest (see text).}
\end{figure}

Here, we present the results of a non-perturbative model which under certain approximations gives analytical expressions for the ADs. Interestingly, these ADs for atomic targets can be interpreted as the ones coming from two-center interferences. In Fig. \ref{fig:illustration}, we present a sketch of the reaction in which an ATPT ionize the atomic target producing several wavepackets that react in the presence of the NIR electric field producing two emitting centers. The interference between the waves emitted from each center will govern the ADs.

Let us consider the photoionization of atomic targets by ATPTs produced by HHG assisted by a monochromatic NIR. The transition matrix amplitude in the dipole approximation in the velocity gauge is given by,

\begin{equation}
M_{SCV}(\mathbf{p})=-i\int_{-\infty}^{\infty}dt\; \langle\Psi_f(\mathbf{r},t)\vert\mathbf{A}(t)\cdot\hat{\mathbf{p}}\vert\Psi_i(\mathbf{r},t)\rangle,
\label{eq:matrix}
\end{equation} where $\mathbf{p}$ is the momentum  of the photoelectron, associated to the momentum operator $\hat{\mathbf{p}}$, and $\Psi_i(\mathbf{r},t)$ and $\Psi_f(\mathbf{r},t)$ are the initial and final wavefunctions, respectively. The vector potential $\mathbf{A}(t)$, representing an ATPT with gaussian envelope, may be written as,
\begin{align}
\mathbf{A}(t)=\mathbf{\Pi}(\phi)\sum_j A_{j} e^{-i j \omega_0 t} e^{i\phi_j} e^{-t^2/2\tau_{T}^2}, \label{eq:pot_vec2}
\end{align}       
where $\omega_0$ is the NIR frequency from which the ATPT is generated, and $\mathbf{\Pi}(\phi)$ represents the polarization vector, respectively. The parameters $\tau_T$ and $\phi_j$ represent the overall duration of the ATPT and the individual phase of each frequency component, whose amplitude is given by $A_j$, respectively. 

The asymptotic final states $\Psi_f(\mathbf{r},t)$ are represented by a spatial wavefunction times a Volkov phase describing the interaction of a free electron of momentum $\mathbf{p}$ with an electromagnetic field  represented through its vector potential $\mathbf{A}_L$.  For the sake of simplicity, we consider a NIR polarized collinearly to the ATPT, and with vector potential given by, 

\begin{equation}
\label{eq:pot_vec_NIR}
\mathbf{A}_L(t)=-\frac{\mathbf{E}_1}{\omega_0}\sin(\omega_0 t-\phi_L),
\end{equation}
where $\mathbf{E}_1$ is the corresponding electric field amplitude and $\phi_L=\omega_0 t_0$ is an arbitrary phase by which the ATPT-NIR delay $t_0$ may be modified.

%\section{The Jacobi-Anger version as usual}
It can be shown \cite{YPCB,Galan2013,Boll2014} that the replacement of Eqs. \eqref{eq:pot_vec2} and \eqref{eq:pot_vec_NIR} into Eq. \eqref{eq:matrix}, leads to the following expression for $M_{SCV}(\mathbf{p})$, 

\begin{align}\begin{split}\label{eq:M_scv}
M_{SCV}(\mathbf{p})&= -i\sqrt{2\pi} \tau_T M_{ph}(\mathbf{p})\sum_{m,n=-\infty}^{\infty} \sum_j i^{n}(-1)^{m} A_j \\
&\times J_{m}(M) J_{n}(N)e^{-i(2m+n)\phi_L} e^{i\phi_j}  e^{-\omega_j^2\tau_T^2/2},
\end{split}
\end{align}
where we have defined, 
\begin{align}
M&=E_{1}^2/(2\omega_0)^3,\label{eq:M1}\\
N&=\mathbf{p}\cdot\mathbf{E}_{1}/\omega_0^2,\label{eq:N12}\\
\omega_j&=p^2/2+I_p+(2M+2m+n-j)\,\omega_0,\label{eq:ome_j}
\end{align}where $p=\vert \mathbf{p} \vert$ and $I_p$ are the asymptotic momentum modulus and the ionization potential of the target, respectively. $M_{ph}(\mathbf{p})$ is the monochromatic transition matrix element describing photoionization by an XUV photon.

Now, if all the frequency components of the ATPT are considered to be of equal strength $A_j$ and also with the same phase $\phi_j$, analytical expressions for the triple sum in Eq. \eqref{eq:M_scv}, corresponding to the AD of photolines satisfying the relation $p^2_q/2+I_p+2M\omega_0=q\omega_0$, can be found. For in-phase odd harmonics ATPTs, the index $j$ runs over odd integer numbers. As the parameter $q$ describing the photoline of interest can be an even or odd integer number, we separate the results according to $q$. For $q$ an even number (SBs) we have, 
\begin{align}
M_{SCV}(\mathbf{p}_q)=i B_0 M_{ph}(\mathbf{p}_q) e^{i M\sin(2\phi_L)} \sin(N\cos\phi_L),\label{eq:M_side}
\end{align} 
whereas for $q$ an odd number we obtain (DHs),
\begin{align}
M_{SCV}(\mathbf{p}_q)= B_0 M_{ph}(\mathbf{p}_q) e^{i M\sin(2\phi_L)} \cos(N\cos\phi_L),\label{eq:M_harm}
\end{align}
where $B_0=-i\sqrt{2\pi} \tau_T A_0$. The constant $M$ appears as a global phase indicating that, independently of the NIR intensity, the oscillatory term that comes from $A_L^2$ in the Volkov phase can be omitted.

Recalling that $N=\mathbf{p}\cdot\mathbf{E}_1/\omega_0^2$, then we define $\mathbf{R}=2\mathbf{E}_1\cos(\phi_L)/\omega_0^2$ and replace it into Eqs. \eqref{eq:M_side} and \eqref{eq:M_harm} we obtain the following expressions for the differential cross sections, 
\begin{align}
    \frac{d\sigma}{d\Omega_e}=\vert B_0\vert^2\,\vert M_{ph}(\mathbf{p}_q)\vert^2\sin^2\left(\mathbf{p}_q\cdot\mathbf{R}/2\right),\label{eq:SB_lines}
\end{align} 
for SB lines and
\begin{align}
    \frac{d\sigma}{d\Omega_e}=\vert B_0\vert^2\,\vert M_{ph}(\mathbf{p}_q)\vert^2\cos^2\left(\mathbf{p}_q\cdot\mathbf{R}/2\right),\label{eq:DH_lines}
\end{align} 
for DH lines, where $\Omega_e$ is the solid angle defining the direction of the photoelectrons.

These results indicate that the ADs for the ionization of an atomic target by a sequence of in-phase odd harmonics in the presence of a NIR reproduce the far-field behavior of two radiating antennas in counter-phase or in phase, respectively, separated a distance $R$. Alternatively, they can be regarded as the ADs for the monochromatic ionization of a homonuclear diatomic molecule oriented collinearly to the NIR \cite{CF,WB,Ciappina2014}, where the DHs (SBs) play the role of the bound-continuum transition of molecules from an initial state of \emph{gerade} (\emph{ungerade}) symmetry.

Interestingly, the distance $\mathbf{R}$ may be interpreted as twice the classical amplitude of a charged particle oscillating in an electric field modulated by the factor $\cos\phi_L$. In this way, the separation $\mathbf{R}$ between the sources can be fixed modifying the ATPT-NIR delay.

To calculate the photoionization AD for atomic targets by using Eqs. \eqref{eq:SB_lines} and \eqref{eq:DH_lines}, we need to compute the corresponding monochromatic transition amplitudes. Provided that the magnetic sublevels of the atomic target are equally populated, the photoionization by linearly polarized monochromatic radiation has the general form \cite{Cooper1968},
\begin{align}\label{eq:cooper}
\vert M_{ph}(\mathbf{p}) \vert ^2 = \frac{\sigma_{tot}}{4\pi} [ 1+\beta P_2(\cos\theta_e) ],
\end{align} where $\sigma_{tot}$ is the total photoionization cross section, $\beta$ is the asymmetry parameter, $P_2(x)$ is the second-order Legendre polynomial and $\theta_e$ is the polar angle between the asymptotic momentum $\mathbf{p}$ and the ATPT polarization vector $\mathbf{\Pi}(\phi)$.  
\par
\begin{figure}[t!]
\includegraphics[width=\columnwidth]{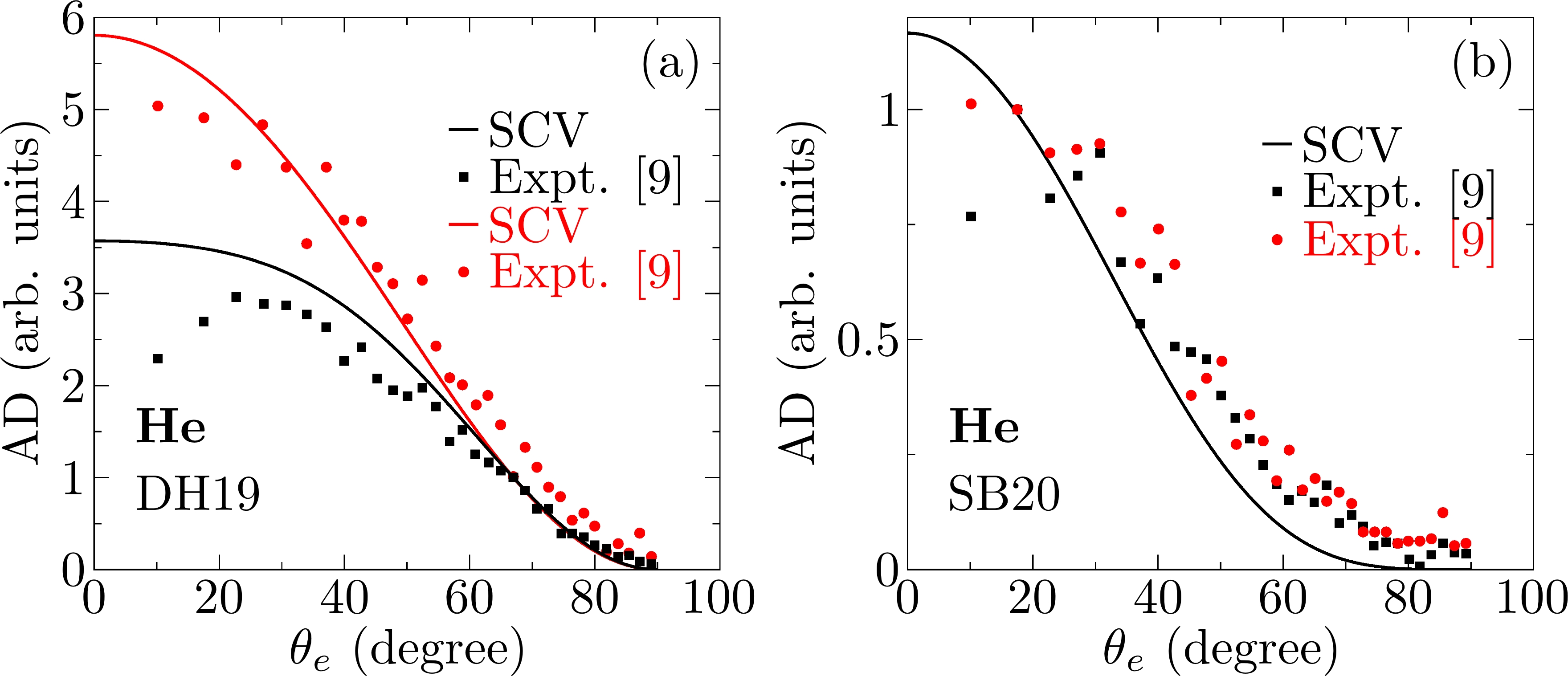}
\caption{\label{fig:He_AD}(Color online) (a) Helium ATPT-NIR angular distributions for the DH19 line with $\phi_L=0.14\pi$ (black) and $\phi_L=0.64\pi$ (red). (b) Same as (a) but for the SB20 line with $\phi_L=0.14\pi$ (black). }
\end{figure}

In Fig. \ref{fig:He_AD}, our analytical results for He targets are compared to the available experimental data \cite{Picard2014}. We consider photoionization by a linearly polarized odd-harmonics ATPT assisted by a collinear NIR of intensity $I_L=0.8\times 10^{12}$ W/cm$^2$, in agreement with the observed shifts due to the ponderomotive energy \cite{Picard2014}. The ADs in Fig. \ref{fig:He_AD} (a) and (b) have been normalized for the emission angles $67^{\circ}$ and $17^{\circ}$, respectively, preserving thus the normalization of the original experimental data \cite{Picard2014}.  

In Fig. \ref{fig:He_AD} (a), the ADs of the DH19 line evolve notably by changing the delay $\phi_L$, particularly for emission in directions close to the polarization axis $(\theta_e=0^{\circ})$. In addition, if $\mathbf{R}$ is smaller than half the wavelength of the photoelectron (as in these ADs), the most sensitive emission direction, \emph{i.e.} the one with the largest path difference, is given by $\theta_e=0^{\circ}$; moreover, total destructive interferences are not expected according to Eq. \eqref{eq:DH_lines} regardless of the emission angle. On the contrary, for emission in the direction $\theta_e=90^{\circ}$ total constructive interferences are expected but the monochromatic AD given by Eq. \eqref{eq:cooper} for $\beta=2$ \cite{Cooper1968} is proportional to $\cos^2\theta_e$, erasing thus the maximum value.

On the other hand, the shape of the AD in the SB$20$ (Fig. \ref{fig:He_AD} (b)) does not show a significant dependence with the delay after normalization. The results for the SBs may be understood taking into account that under the present conditions, the SBs are populated mainly by two-photon transitions. So, an expansion of Eq. \eqref{eq:M_side} into Bessel functions retaining only the first term leads to,  
\begin{align}   
\frac{d\sigma}{d\Omega}=\vert B_0 \vert^2 \vert J_1(N) \vert^2 \vert M_{ph}(\mathbf{p}) \vert ^2 \cos^2 \phi_L\label{eq:SB_simple}
\end{align}
where the ATPT-NIR delay dependence turns out to be a scaling factor. This result is similar to  the one found in previous second order perturbation studies for SBs \cite{Paul2001,Picard2014}. The faster decay observed, for angles $\theta_e$ near $90^{\circ}$, in the SB's angular distribution as compared to the DH ones may be explained as the superposition of the monochromatic $\cos^2\theta_e$ decay, modulated, in the SBs, by the destructive interference predicted by Eq. \eqref{eq:SB_lines} for $\theta_e=90^{\circ}$. The slight discrepancies observed between the theoretical predictions and the experimental data for emission angles near $90^{\circ}$ are likely to result from our model that forbids the exchange of NIR photons for emission in directions perpendicular to the NIR polarization one. In contrast, the TDSE results show that the ADs of the SBs do no cancel completely at $90^{\circ}$ \cite{Guyetand2005}.
     
\begin{figure}[t!]
\includegraphics[width=\columnwidth]{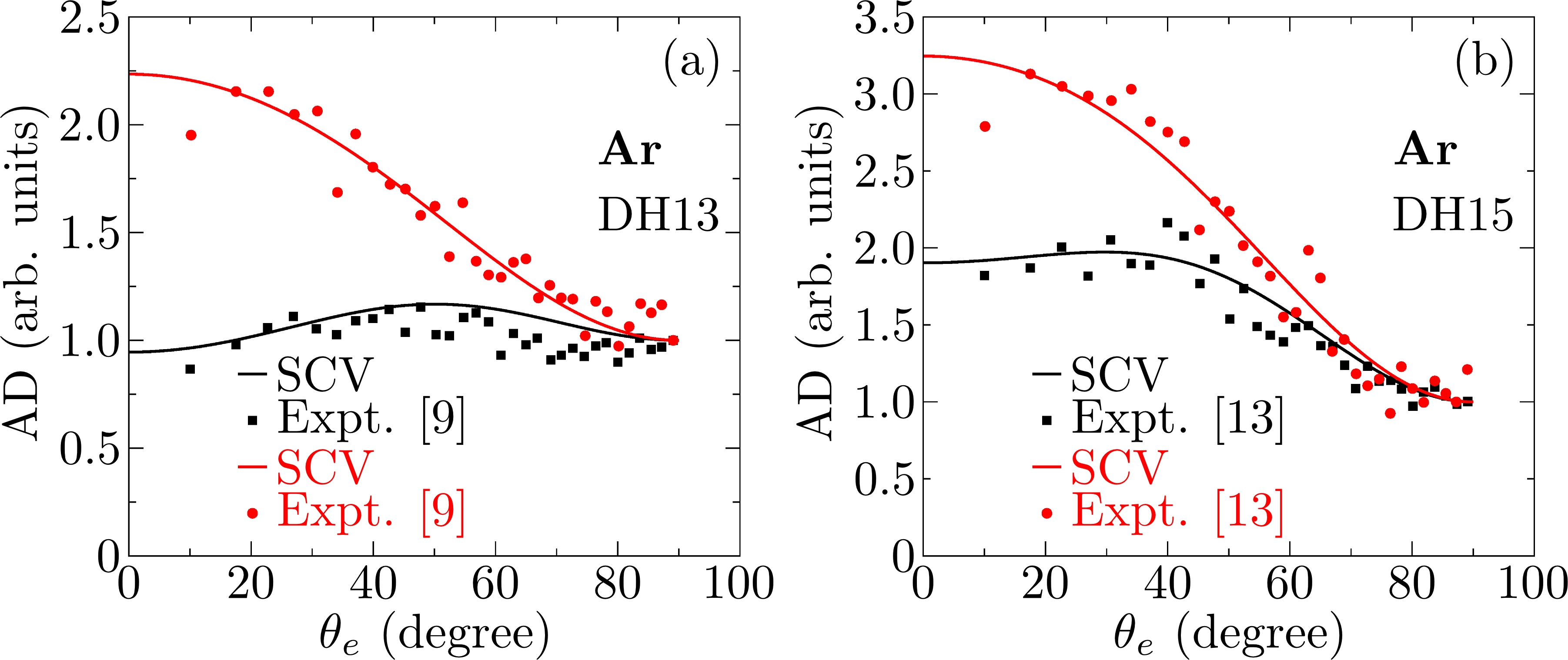}
\caption{\label{fig:AR_AD}(Color online) (a) Argon ATPT-NIR angular distributions for the DH13 line with $\phi_L=0.144\pi$ (black) and $\phi_L=0.644\pi$ (red). (b) Same as (a) but for the DH15 line with $\phi_L=0.188\pi$ (black) and $\phi_L=0.688\pi$ (red). }
\end{figure} 
In Fig. \ref{fig:AR_AD}, we show ADs for the DHs of Ar targets, obtained with our model and compared to the available experimental data \cite{Picard2014,Weber2015}. These results are in better agreement with the experiments as it is the case also for the soft-photon approximation \cite{Picard2014}. As in the He case, we consider a linearly polarized in-phase odd-harmonics ATPT assisted by a collinear NIR of intensity $I_L=1.3\times 10^{12}$ W/cm$^2$ for Fig. \ref{fig:AR_AD} (a) and $I_L=0.76\times 10^{12}$ W/cm$^2$ for Fig. \ref{fig:AR_AD} (b). The normalization of the ADs has been applied for an emission angle of $90^{\circ}$. The asymmetry factor $\beta$ corresponding to each of the asymptotic photoelectron energies were interpolated from theoretical data from \cite{Huang1981}. For this target, as the emission in the direction perpendicular to the ATPT polarization  does not cancel because of the dipole selection rules \cite{Cooper1968}, then constructive interferences predicted by Eq. \eqref{eq:DH_lines} for emission at $\theta_e=90^{\circ}$ in the DH lines may be observed. The evolution of the ADs as a function of the ATPT-NIR delay for emission angles near $\theta_e=0^{\circ}$ may be interpreted analogously to the He case.   
\par
So far, we proved that our model is reasonably accurate in describing the ADs in several situations. Now, we proceed a step further by considering a higher NIR intensity and photoelectron energies satisfying the conditions for the existence of nodes in DH lines coming from two-center interferences, i.e., $R>\lambda_q/2$ (or $\mathbf{p}_q\cdot \mathbf{R}>\pi$). 

\begin{figure}[t!]
\includegraphics[width=\columnwidth]{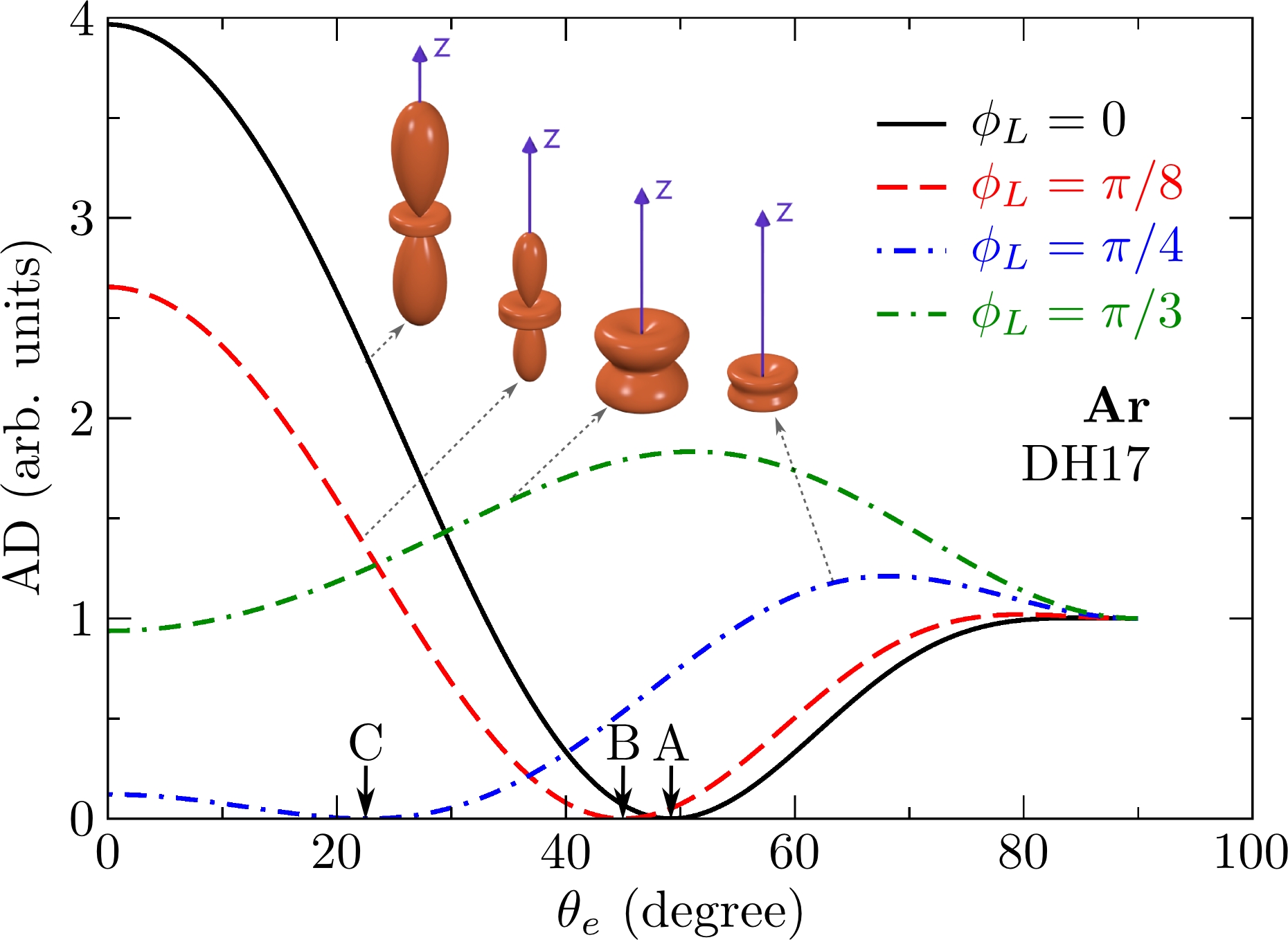}
\caption{\label{fig:AR_AD_+I_H17}(Color online)  Argon ATPT-NIR angular distributions for the DH$17$ line for different delays $\phi_L$.}
\end{figure}

In Fig. \ref{fig:AR_AD_+I_H17}, we have simulated the ATPT-NIR angular distributions for the DH$17$ in Ar for different delays $\phi_L$. The ATPT contains only odd harmonics and the intensity of the NIR is chosen to satisfy the relation $N_{max}=2.4048$, \emph{i.e.}, the first zero for $J_0(z)$. The NIR intensity satisfying the above conditions is $I_L=2.78\times 10^{12}$ W/cm$^2$, that is about two or three times the intensity of the previous cases. These ADs share two features, namely, for $\theta_e=0$ no one-photon contribution is present and for $\theta_e=\pi/2$ only the one-photon contribution is present, according to the prescriptions of the SCV model.

The angles labeled as A, B and C in Fig. \ref{fig:AR_AD_+I_H17} satisfy the relation $\cos(\mathbf{p}_q\cdot \mathbf{R}/2)=0$, showing that with a higher NIR intensity and/or photoelectron energy, it could be possible to observe nodes in the ADs. Moreover, these zeros have been found earlier in \emph{ab-initio} calculations (Fig. (5) from Ref. \cite{Galan2013}). Also, it is clear that  as the delay $\phi_L$ increases, the angular position of these zeros moves towards $\theta_e=0^{\circ}$, \emph{i.e.}, as the separation between the emitters is reduced, the photoelectron loses its capacity to scan the two-center structure and finally for $\phi_L=\pi/3$ no destructive interferences are observed because the separation $R$ becomes smaller than half the photoelectron wavelength.   

\begin{figure}[t!]
\includegraphics[width=\columnwidth]{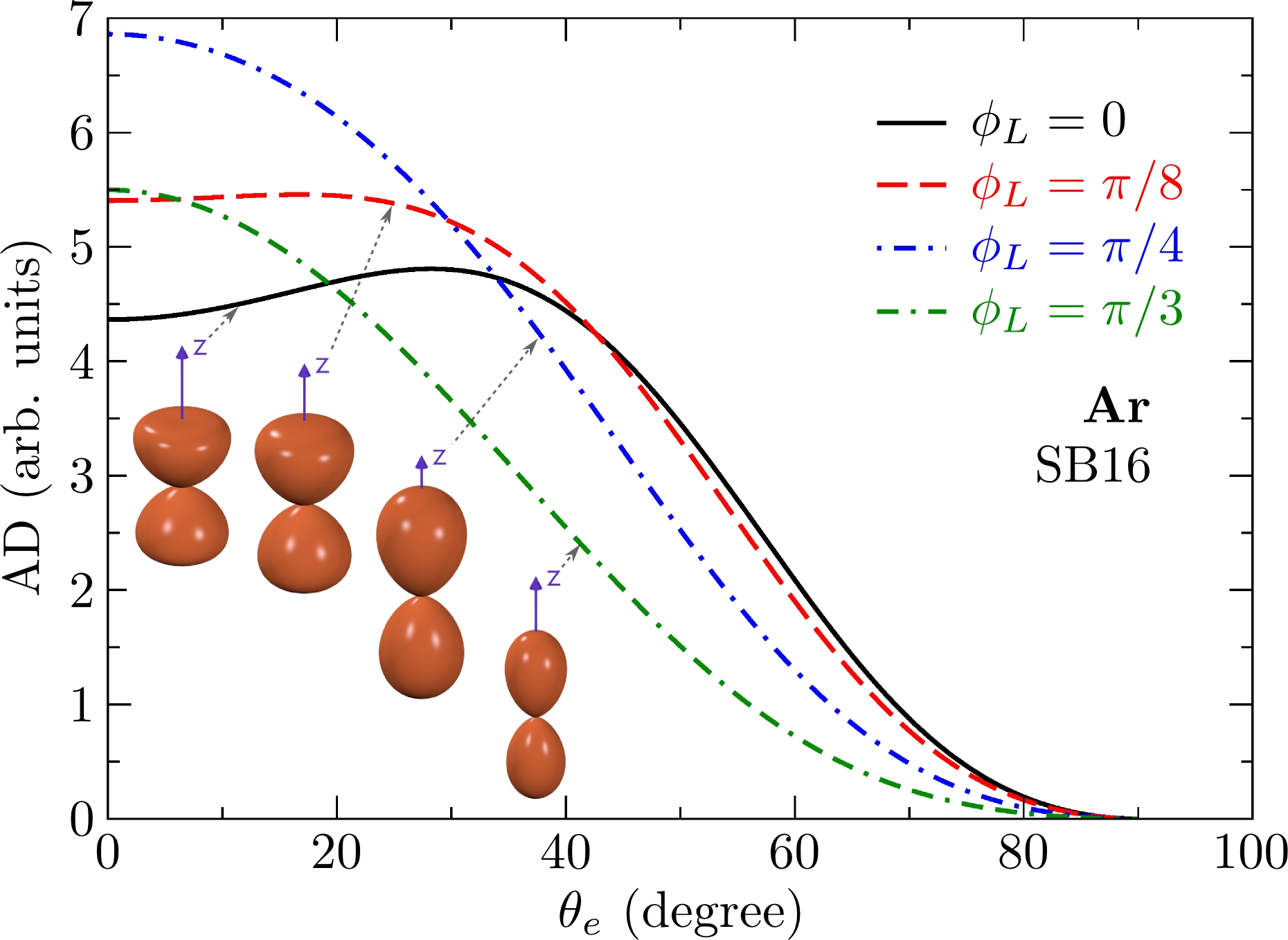}
\caption{\label{fig:AR_AD_+I_SB16}(Color online) Argon ATPT-NIR angular distributions for the SB$16$ line for different delays $\phi_L$.}
\end{figure}

The results corresponding to the SB16 in Fig. \ref{fig:AR_AD_+I_SB16} are different from those of Helium in Fig. \ref{fig:He_AD} (b) mainly because the ADs now show an evolution when different delays are considered. This indicates that the Eq. \eqref{eq:SB_simple} is no longer appropriate to describe the ADs of sideband lines.

In contrast with the DH lines, the SB lines do not show nodes in the ADs for the considered NIR intensity and photoelectron energy. Moreover, this may be understood recalling the condition for the existence of nodes in the SBs according to Eq. \eqref{eq:SB_lines}, $\mathbf{p}_q\cdot \mathbf{R}>2\pi$, requiring thus twice the NIR vector potential amplitude or twice the photoelectron momentum compared to the dressed harmonic lines case. Furthermore, for a given delay, the expected correlation between the increase (decrease) in the DH lines with a decrease (increase) in the SB lines is observed.

In summary, we have modified the SCV model to address the laser assisted photoionization of atomic targets by ATPTs arising from high harmonic generation. Even if the double-slit behavior was inferred previously for experiments in the streaking regime, the closed-form results obtained with our model show clearly the two-center nature of the interfering wave packets, providing simple expressions for the ADs of photoelectrons. Moreover, these analytical expressions might be useful to achieve the attosecond control of the electron dynamics in the  intermediate NIR intensity range. Additionally, they may be employed to extract information about the NIR or the delay from the position of the zeros in the ADs.   

 In turn, the overall good agreement with the experimental results is encouraging as our calculations may be extended without great effort to more complex targets or ATPTs, like those leading to parity mix interferences. Work in this direction is in progress.

Authors acknowledge financial support from the Agencia Nacional de Promoci\'on Cient\'ifica y Tecnol\'ogica (PICT No. 01912), the Consejo Nacional de Investigaciones Cient\'ificas y T\'ecnicas de la Rep\'ublica Argentina (PIP No. 11220090101026), and the Fundaci\'on Josefina Prats.

\bibliography{report}

\end{document}